\documentclass[11pt]{article}
\usepackage{latexsym,cite,amssymb}
\usepackage{times}
\makeatletter

\@addtoreset{equation}{section} \makeatother

\input amssym.def
\input amssym.tex

\newcommand{\G}{{\Gamma}}
\newcommand{\g}{{\gamma}}

\newcommand{\be}{\begin{equation} }
\newcommand{\ee}{\end{equation} }
\newcommand{\ba}{\begin{array}}
\newcommand{\ea}{\end{array}}

\newcommand{\su}{\mbox{su}}
\newcommand{\so}{\mbox{so}}

\def\I_N{{1_{\scriptscriptstyle N\times N}}}

\def\sI{{I}}
\def\sJ{{J}}

\newcommand{\nn}{\nonumber}

\def\g{\gamma}
\def\G{\Gamma}

\def\ep{\epsilon}

\def\m{\mu}

\def\s{\sigma}

\begin{document}
\begin{titlepage}
\title{\vskip -60pt
\vskip 20pt Abelian Vortices in Bagger-Lambert-Gustavsson Theory\\
~}
\author{\sc Jongwook Kim 
 and \sc Bum-Hoon Lee
}
\date{}
\maketitle \vspace{-1.0cm}
\begin{center}
~~~
Department of Physics / Center for Quantum Spacetime, Sogang University , Seoul 121-742, Korea\\
~{}\\
{\small{Electronic correspondence: {{{jongwook@sogang.ac.kr}}}}}
~~~\\
~~~\\
\end{center}
\begin{abstract}
\noindent

Among newly discovered $M2$ and $M5$ objects in the Bagger-Lambert-Gustavsson theory, our interest is about ${1\over 2} $ BPS vortices, which are described by covariantly holomorphic curves in transverse coordinates.  We restrict ourselves to the case where the global symmetry is broken to $\so(2) \times \so(2)\times \so(4)$ for the mass deformed Bagger-Lambert theory.
 A localized object with finite energy exists in this theory since the mass parameter supports regularity. It is time independent but carries angular momentum coming solely from the gauge potential by which the energy is bounded below.
\end{abstract}
{\small
\begin{flushleft}
\end{flushleft}}
\end{titlepage}
\newpage

\tableofcontents 
\section{Introduction}
The Bagger-Lambert-Gustavsson(BLG)theory \cite{BL} \cite{Gustavsson} is a three dimensional
Chern-Simons-Higgs system with superconformal invariance for M2 branes. It has 16 supersymmetries and $\so(8)$ global symmetry. The theory is based on the gauge symmetry generated by three algebra, and it is known that $\so(4)$ is the unique choice for its realization of the finite representation with ghost free theory\cite{3algebra}. The generalization beyond $\so(4)$ has been studied using Lorentzian representation\cite{Lorentzian} or infinite dimensional representation \cite{Nambu}, and the latter has the natural connection to M5 physics. The formulation of $N$ number of interacting $M2$s in flat space time has been proposed  by Aharony-Bergman-Jafferis-Maldacena \cite{ABJM}(ABJM). It is $\mathcal N = 6$ $U(N)\times U(N)$ Chern-Simons gauge theory with $\su(4)$ global symmetry with matter interaction given by a quartic superpotential. Still the realization of $\mathcal N = 8$ theory is not known yet. Another way of generalization  both for BLG\cite{massdeformBL} and ABJM \cite{HL3P}\cite{HL3PN=56}\cite{massivestudyM2} can be performed by mass deformation, which breaks scale invariance and the global symmetry while keeping the supersymmetries.

The BLG theory and the ABJM theory have interacting M2-brane description and also the theory portray supersymmetric objects of the 11 dimensional quantum gravity. Those object will be obtained from classical BPS solutions in the dual field theory. Some of ${1\over 2}$ BPS equations have been written and also the solutions have been studied by the various authors. The systematic classification of BPS objects has been done in \cite{OctonionBPSBL} and some of their solutions have been further studied in \cite{NWKIM}. New $M2$, $M5$ objects are actively investigated in  Bagger-Lambert theory \cite{BL} \cite{Gustavsson} \cite{Lorentzian} \cite{Nambu} \cite{massdeformBL} \cite{NWKIM} \cite {Filippo} \cite{Furuuchi}  \cite{BasuHarvey} and in $\mathcal N=6 $ Chern-Simons gauge theory of ABJM \cite{ABJM} \cite{HL3P}\cite{HL3PN=56} \cite{massivestudyM2} \cite{BL:N=6} \cite{BandresLIpsteinSchwartz} \cite{Raamsdonk} \cite{HanakiLin} \cite{Alessandro} \cite{Carlo.Maccaferri}. It is found that there are various objects like fuzzy funnels, fuzzy spheres, domain walls, and vortices.  Our interest is about vortices, which are covariantly holomorphic curves in the transverse coordinates in terms of membrane perspective. 
 We search for such an object by assimilating vortices of 2+1 dimensional Chern-Simons-Higgs theory in the mass deformed Bagger-Lambert theory \cite{massdeformBL} when $\so(4)\times \so(4)$ global symmetry is broken to $\so(2) \times \so(2)\times \so(4)$. We adopt $\so(4)$ representation of the BLG theory, which is equivalent to $\mathcal N = 8$ Chern-Simons-Higgs theory with $\su(2)\times \su(2)$ gauge symmetry \cite{Raamsdonk}. It may be convenient to take up the latter representation to generalize the result  into ABJM theory, but the three bracket notation in $\so(4)$ representation provides simpler and clearer way to construct our problem.  
 
The size of the vortex is proportional to the inverse square root of the mass parameter and its energy is bounded below by quantized magnetic flux in the Chern-Simons-Higgs theory. For Abelian $\mathcal N =1$ or $\mathcal N = 2$ Chern-Simons-Higgs theory, mass parameter of the theory must be introduced to have a regularized finite energy vortex configuration \cite{abelianvortexCS} \cite{abelianvortexCS:Yoonbai}. For Maxwell-Higgs theory Fayet-Iliopoulos(FI)-parameter play the same role \cite{MirrorSymmetry} to regularize the vortex configuration. We have only singular extended and infinte energy object in the Bagger-Lamber theory \cite{NWKIM}, but we have regular localized and finite energy object in the mass deformed Bagger-Lambert theory, since the mass parameter resolves regularity issues.

In the following main section, we start with the mass deformed Bagger-Lamber action\cite{massdeformBL}. Then we derive a set of $1\over 2$ BPS equations and check the energy of the system is bounded below by central charges. In the first subsection (\ref{sectionvortexeqs}) we show how the vortex equations are brought about and how regularity issues are resolved. In the second subsection (\ref{sectionvortexsolution}) finally we provide the explicit solution whose energy is bounded below by quantized charges.

\section{Half BPS configuration in the mass deformed Bagger-Lambert theory}

We start our discussion by invoking the bosonic part of the mass deformed Bagger-Lambert theory. It is \cite{massdeformBL}
\begin{eqnarray}
\mathcal L &=& -{1\over 2} D_\mu X_I D^\mu X_I  - V (X^I)+ {1 \over 2} \epsilon^{\mu \nu \lambda} (f^{abcd} A_\mu{}_{~ab} \partial_\nu A_\lambda{}_{~cd}
                + {2 \over 3} f^{cda}{}_g f^{efgb} A_\mu{}_{~ab} A_\nu{}_{~cd} A_\lambda{}_{~ef})\, , \nn \\
 \end{eqnarray}
where
 \begin{eqnarray}   
     V(X^I) &=& {1\over 2\cdot 3!} [X_I, X_J, X_K][X_I, X_J, X_K] + {m^2 \over 2} X_I X_I + 4m (X_1 [X_2, X_3, X_4] - X_5 [X_6, X_7, X_8]) \,, \nn
     \\ \label{potential0}
\label{massdeformedaction}
\end{eqnarray}
with $I=1,2,\cdots,8$ and $\mu= t, x, y$. Here $a,b,c,d$ are gauge indices.\footnote{We restrict ourselves to Chern-Simon's level $\kappa = - 1$. In \cite{massdeformBL}, the authors use different spinor conventions  $\Gamma^{12345678} \epsilon = -\epsilon$ in contrast to ours such that  $\Gamma^{12345678} \epsilon = \epsilon$. It differs by labeling 7 and 8 in exchange, so we have sign difference in the quartic term.} $D_{\mu}X^{I}_{a} =\partial_{\mu}X^{I}_{a}-\tilde{A}^{b}{}_{a}X^{I}_{b}$ and $\tilde A_\m{}^a{}_b = A_\mu{}_{cd} f^{cda}{}_bt$. Three bracket is defined as $[X_I, X_J, X_K]_a = f_{abcd} X_I{}^b X_J{}^c X_K{}^d$ where we adopt $\so(4)$ representation $f^{abc}{}_{d}=f^{abcd}=\varepsilon^{abcd}$ \cite{BL}.
We see that $\so(8)$ global symmetry is broken to $\so(4)\times \so(4)$ by the mass term.
All the variables are three-algebra valued. The trace over whole expression is implied and will be omitted when its meaning is obvious. Recall that in contrast to the original convention \cite{BL} we take $\mu\equiv 0,9,{10}$ directions as for the M2-brane  worldvolume for simplicity\cite{OctonionBPSBL},
\be
\ba{ccc}
x^{0}{\equiv t}\,,~~~~&~~~~x^{9}{\equiv x}\,,~~~~&~~~~x^{10}{\equiv y}\,.
\ea
\ee
The energy momentum tensor is
\be\ba{ll}
T_{\mu \nu} = D_\mu X^I _{~a} D_\nu X^{a I}  - \eta_{\mu \nu} ({1\over 2} D_\rho X^{a I} D^{\rho } X_a^{~I} + V(X^I) )\,.
\ea\ee
The Chern-Simons term doesn't contribute to the energy momentum because it is topological. The Hamiltonian is readily obtained from $T_{00}$. There are several choices of completing squares of the Hamiltonian according to BPS configurations we are interested in. Considerable number of BPS equations were classified according to global symmetries on $M2$ \cite{OctonionBPSBL}. For our pursuit to the vortex configuration among various $M2$ intersecting on $M2$s, we concentrate on the sector where $\so(2) \times \so(2)\times \so(4)$ global symmetry exists. Supersymmetry transformation for the spinors are
\be
\delta\Psi=\left(F_{\mu I}\Gamma^{\mu I} + {m } \Gamma^{1234} X_I^a \Gamma_I  -  \textstyle{\frac{1}{6}}F_{IJK}\Gamma^{IJK}\right)\epsilon\,,
\label{BPS0}
\ee
where all the variables are three-algebra valued again and we set
\be
\ba{ll}
F_{\mu I}\equiv D_{\mu} X_{I}\,,~~~~&~~~~~F_{IJK}\equiv\left[X_{I},X_{J},X_{K}\right]\,.
\ea
\ee
By imposing the BPS condition in this transformation (\ref{BPS0}), we derive the BPS equations. 
The mass deformed theory has $\so(1,2) \times \so(4) \times \so(4)$ global symmetry. BPS equations for q-balls and vortices in the selfdual Chern-Simons Higgs theory were shown to describe the $1\over 4$ BPS configuration in mass deformed BLG theory \cite{massdeformBL} with two projectors as ${1\over 2} (1 \pm \Gamma^{xy12})\mathcal P$ and ${1\over 2} (1 \pm \Gamma^{xy34})\mathcal P$ \cite{massdeformBL}, where $\cal P $ $= {1\over 2} (1 + \Gamma^{12345678})$.  We search for a  $U(1)$ vortex imbedding in the mass deformed Bagger-Lambert theory with $\so(4)$ gauge group in the $1\over 2$ BPS configuration. We choose projector for supersymmetry parameter as $ {1\over 2} (1 + \Gamma^{xy12}) \cal P $ to project out $\cal N$ = 8 supersymmeties leaving $\so(2) \times \so(2) \times \so(6)$ isometry and in effect the BPS configuration should have $\so(2) \times \so(2) \times \so(2) \times \so(4)$ global symmetry. BPS equations of mass deformed Bagger-Lambert theory are read from
\begin{eqnarray}
\Big(F_{\m I}^{~~a}\G^{\m I} + {m} \Gamma^{1234} X_I^a \Gamma_I  - {1 \over 6} F_{IJK} \G^{IJK} \Big)  (1 + \Gamma^{xy12}) \mathcal{ P} \ep = 0
\end{eqnarray}

We take the eleven-dimensional gamma matrix representation that makes the $\so(8)$ symmetry incarnate. The eleven-dimensional $32{\times 32}$ gamma matrices $\Gamma^{M}$, $M=\mu,I$, $\mu=t,x,y$, $I=1,2,\cdots,8$ in   Bagger-Lambert theory naturally decompose into two parts: $\so(1,2)$ for the M2-brane worldvolume and  $\so(8)$ for the transverse space
\be
\ba{llll}
\Gamma^{t}=\varepsilon\otimes\gamma_{(9)}\,,~~&~~~
\Gamma^{x}=\sigma_{1}\otimes\gamma_{(9)}\,,~~&~~~
\Gamma^{y}=\sigma_{3}\otimes\gamma_{(9)}\,,~~&~~~
\Gamma^{I}=1\otimes\gamma^{I}\,,~~~~~~I=1,2,\cdots,8\,.
\ea
\ee
Gamma matrices for $\so(1,2)$ were chosen as $\s^\mu=\{\varepsilon,\s^1,\s^3\}$ all of which are real. Then 11 dimensional Majorana condition is trivial when we pick representation for $\so(8)$ gamma matrices to be real too.
Here $\gamma^{I}$'s are the $16{\times 16}$  gamma matrices in the  eight-dimensional Euclidean space and $\gamma_{(9)}\equiv\gamma_{12\cdots8}$. Clearly in this representation the chirality of $\so(1,2)$  coincides with that of $\so(8)$
\be
\Gamma^{txy}\epsilon = \Gamma^{12345678} \epsilon = \epsilon \,.
\ee
In addition,
\be
\ba{lll}
\gamma_{\sI}=\left(\ba{cc}0~&~\rho_{\sI}\\{}&{}\\(\rho_{\sI})^{T}~&~0\ea\right)\,,~~&~~~
\rho_{\sI}(\rho_{\sJ})^{T}+\rho_{\sJ}(\rho_{\sI})^{T}=2{\delta_{\sI\sJ}}\,,
~~&~~~\gamma_{(9)}=\gamma_{12345678}=\left(\ba{rr}1&0\\0&-1\ea\right)\, .
\ea
\ee
Here $\Gamma^{txy} = 1\otimes\gamma_{(9)}$. It is consistent with the fact that the product of all the eleven-dimensional gamma matrices leads to the identity $\Gamma^{txy123\cdots 8}=1$.
It is convenient to decompose $32 \times 32$ gamma matrices under chirality condition $\G^{txy}\ep_{[32]} = \ep_{[32]}$, meaning identically $ \g^{(9)} \ep_{[16]}^i = \ep_{[16]}^i$ when $\ep_{[32]} = \pmatrix{\ep^1_{[16]} \cr \ep^2_{[16]}}$. Our projector in this representation is
 \begin{eqnarray}
(1 + \Gamma^{xy12}) \mathcal P = \pmatrix{1 & 0 & -\rho^1{}^T \rho^2 & 0\cr
                                        0 & 0 & 0 & 0 \cr
                                        \rho^1 \rho^2{}^T & 0 & 1 & 0 \cr
                                        0&0&0&0} \,.
\end{eqnarray}
After a routine calculation of arranging gamma matrix products and reading coefficients in front of them and setting them to vanish, we obtain the set of ${1\over 2}$ BPS equations as,
\begin{eqnarray}
  & F_{x1} + F_{y2} = 0 \,, ~~~~ F_{x2} - F_{y1} = 0 \,, \label{holomorphiccovariance}\\
 &F_{t 1} = 0\,,~~~F_{t 2} = 0\,, ~~F_{t3} + F_{312} - m X_4 = 0 \,,~~~~ F_{t4} + F_{412} + m X_3 = 0 \,~~ ;~~ F_{t \tilde A} = F_{\tilde A 1 2} \,,\label{timedependent}\\
& F_{x3} = F_{y3} = 0 \,,~~~~ F_{x4} = F_{y4} = 0 \, ~~; ~~F_{x \tilde A} = F_{y \tilde A} =0\,, \label{covariantlyconstant}\\
& m X_{1} = -F_{234}\,,  ~~m X_2 =  F_{134}\,  ~~;~~ m X_{\tilde A} =  \epsilon_{\tilde A \tilde B \tilde C \tilde D} F_{\tilde B \tilde C \tilde D}\,, ~~~~ \mbox{other } F_{IJK}\mbox{s}\mbox{ vanish.}\label{threealgebrarelation}
\label{massdeformBPSeqns}
\end{eqnarray}
Here $A=1,2,3,4$ and $\tilde A = 5,6,7,8$. In addition, we have to solve the equation of motion for the gauge field. It is
\be
\tilde{F}_{\mu\nu}{}^{a}{}_{b}+\varepsilon_{\mu\nu\lambda}X^{J}_{c}D^{\lambda}X^{J}_{d}f^{cda}{}_{b}=0\,,
\ee
where $\tilde{F}_{\mu \nu}{}^{a}{}_{b} = \partial_\nu \tilde{A}_{\mu}{}^{a}{}_{b} -\partial_\mu \tilde{A}_{\nu}{}^{a}{}_{b} -
\tilde{A}_{\mu}{}^{a}{}_{c}\tilde{A}_{\nu}{}^{c}{}_{b} +\tilde{A}_{\nu}{}^{a}{}_{c}\tilde{A}_{\mu}{}^{c}{}_{b} $\, and $\varepsilon^{txy}=1$ .
Part of the transverse coordinates $X^1$ and $X^2$ together with the $x, y$ of the M2-brane world volume on $SO(1,2)$ are complexified in such a way as
\be
\ba{ll}
X_{{\omega}}= \textstyle{\frac{1}{\sqrt{2}}}(X_{1}-i X_{2})\,, ~&~ D_z =\textstyle{\frac{1}{\sqrt{2}}} (D_x -iD_{y})\,. \label{complexify}
\ea
\ee
The equations (\ref{holomorphiccovariance}) give rise to vorticity and they can be written compactly as,
\be\ba{ll} D_z X_{\bar{\omega}}=0 \,. \label{holomorphiccovarianceXw}\ea\ee
With its complex conjugate we call both as holomorphic covariance. Time dependent solutions are generically existent. $D_t X^I\neq 0$ implies a nonzero momentum
along $X^I$ direction and it is the M-wave. The last equation of (\ref{timedependent}) shows that M-wave momenta along $\so(4)$ are proportional  to non-vanishing three brackets $F_{\bar A \bar B \bar C}$ which are related to the existence of M5-branes. The remaining equations in ( \ref{timedependent} ) show that M-wave momenta along $X_3$ and $X_4$ are proportional to the mass terms as well as three products, and M-wave on holomorphic directions vanishes. Rewriting the Hamiltonian as sum of complete squares, we have
\be\ba{ll}
 H = {1 \over 2} \int dx dy  &
    ~~~( ~| D_t X_1 |^2 + | D_t X_2 |^2 + | D_{x} X_1 + D_y X_2 |^2 + | D_{x} X_2 - D_y X_1 |^2
  \\
  &   + |m X_1 + [X_2, X_3, X_4]|^2 + |m X_2 - [X_1, X_3, X_4]|^2
  \\
   & + |D_x X_3 |^2 + |D_y X_3 |^2 +|D_t X_3 + [X_3, X_1, X_2] - m X_4|^2
  \\
   & + |D_x X_4 |^2 + |D_y X_4 |^2 +|D_t X_4 + [X_4, X_1, X_2] + m X_3|^2
  \\
 & + |m X_{\bar A}
        - \epsilon_{\bar A \bar B \bar C \bar D} [X_{\bar B}, X_{\bar C}, X_{\bar D}]|^2
        + |D_x X_{\bar A}|^2+ |D_x X_{\bar A}|^2
        +|D_t X_{\bar A} + [X_{\bar A}, X_1, X_2]|^2
  \\
  & + [X_1, X_3, X_{\bar A}]^2 + [X_2, X_3, X_{\bar A}]^2 + [X_1, X_4, X_{\bar A}]^2 + [X_2, X_4, X_{\bar A}]^2
  \\
   & + {1 \over 2} [X_A, X_{\bar A}, X_{\bar B}]^2  ~)
   \\
   &
  \\
 & +  \int dx dy ~~(~ Z^{[12]} + R^{[34]} ~)\,. \label{energybound}
\ea\ee
In completing squares the Hamiltonian we again have the equivalent set of BPS equations (\ref{holomorphiccovariance}), (\ref{timedependent}), (\ref {covariantlyconstant} ), (\ref{threealgebrarelation}) that are  obtained from solving the killing spinor equations. The gauss constraint is necessary for verification. The energy is bounded below by two central terms $Z^{[12]}$ and $R^{[34]}$ whose generic definitions are \be\ba{ll}
Z^{IJ} =- \partial_i (X^I D_j X^J \epsilon^{ij})\,, ~~~R^{IJ}&= -2 m (X^{I}D_{0}X^{J}-X^{J}D_{0}X^{I})\,. 
\ea
\ee
 The former $Z^{[12]}$ is the same central charge that appears in \cite{Filippo}, \cite{Furuuchi} and it simplifies further to $ \int d^2 x ~ Z_{12}= \int dz d\bar z \partial_z \partial_{\bar z} ~ (X_\omega^a X_{\bar \omega}{}_a)$ using BPS equations (\ref{holomorphiccovarianceXw}). The latter is nothing but an angular momentum on $X_3$ and $X_4$ plane \cite{massdeformBL}.

\subsection{Vortex} \label{sectionvortexeqs}
For simplicity, we turn off $ X_5, X_6, X_7, X_8 $ consistently with  the equations involving $X_{\tilde A}$s which are at the right sides to the semicolons in (\ref{timedependent}), (\ref{covariantlyconstant}), (\ref{threealgebrarelation}). Then the potential (\ref{potential0}) simplifies :
\be\ba{ll}
V= & {1\over 2}
\left( [X_3,X_1, X_2]^2 + [X_4, X_1, X_2]^2 + [X_1, X_3, X_4]^2 + [X_2, X_3,X_4 ]^2  \right)
    \\ \label{potential}
     &   + { m^2 \over 2} (X_A X_A) + 4m X_1 [X_2, X_3, X_4] \,. \ea\ee
We set $X_1, X_2, X_3, X_4$ as
\be
 \ba{lll}
X_1 = \left(-{\det M \over m }a,{\det M \over m }b~~,0~~~~~~~~,0~~~~~~   \right) \,,& \\
X_2 = \left(b~~~~~~~~~~~~~,a~~~~~~~~~~,0~~~~~~~~,0~~~~~~\right)  \,,& \\
X_3 = \left(0~~~~~~~~~~~~,0~~~~~~~~~~~, M_3{}^3~~,M_3{}^4~  \right) \,, &  \\
X_4 = \left(0~~~~~~~~~~~~,0~~~~~~~~~~~,M_4{}^3~~,M_4{}^4 ~ \right) \,. &
\ea \ee
Note that this ansatz readily solves the first two BPS equations in (\ref{threealgebrarelation}). We will denote $ X_3 $ and $ X_4 $ values depicted in the lower right block-diagonal $2\times 2$ matrix as $M_i{}^j $. From the equation (\ref{holomorphiccovariance}) $A_{z}{}^1{}_3$ or $A_{z}{}^1{}_4$ have trivial values according to the above ansatz therefore,
\be\ba{ll}
\tilde{A}_{z}=\left(\ba{cccc}0&A_{z}{}^1{}_2&0&0\\ -A_{z}{}^1{}_2&0&0&0\\
 0&0&0&A_{z}{}^3{}_4\\0&0&-A_{z}{}^3{}_4&0 \ea\right)\,,&~ \tilde{A}_{t}=\left(\ba{cccc}0&A_{t}{}^1{}_2&0&0\\ -A_{t}{}^1{}_2&0&0&0\\
 0&0&0&A_{t}{}^3{}_4\\0&0&-A_{t}{}^3{}_4&0\ea \right)\,. \label{gaugeansatz}
 \ea
 \ee
Under these ansatz the Gauss constraint is
 \be
\ba{lll}
\tilde{F}_{xy}{}^a{}_b =  - 2{ \det M \over m} \sum_{i,j = 3,4} |M_i{}^j|^2 ( \Phi \bar \Phi - {m^2 \over \sum  |M_i{}^j|^2 })\epsilon^{34a}{}_b\, ,  \label{massdeformgauss}
\ea
\ee
 and the angular momentum is
\be\ba{lll}
R_{34} = 2 m^2  \sum |M_i{}^j|^2 - 4 \det M^2 (a^2+ b^2) \, .
\ea\ee
To have vorticity in convenient way we fix the scale of $ X_3 $ and $ X_4 $ as,
\be\ba {lll}
\det M = m \,. \label{simplicityM}
\ea\ee
 In complex coordinates $X_1$ and 
$X_2$ are written as $X_w = (\Phi, i \Phi , 0, 0 )$ and $X_{\bar w} = (\bar \Phi , -i \bar \Phi, 0, 0 )$ where $\Phi = -{1\over \sqrt 2}(a + i b)$ so that $2 \Phi \bar \Phi = a^2 + b^2$. This means that we restrict ourselves only to the $\alpha(z,t) = 0$ sector in \cite{NWKIM}. Then covariantly holomorphic conditions (\ref{holomorphiccovarianceXw})
become
\be\ba{ll}
D_z \bar \Phi = 0 \, , \label{vortexeqn}
\ea\ee
where $D_z = \partial_z + A_z{}^1{}_2$.
Recall that the gauge transformation of the Bagger-Lambert theory is
\begin{eqnarray}
\delta \tilde{A}_\mu{}^{a}{}_{b} = \partial_\mu \Lambda^a{}_{b} - \Lambda^b{}_{c}\tilde{A}_\mu{}^{c}{}_{a} + \tilde{A}_\mu{}^{b}{}_{c} \Lambda^c{}_{a} \, .
\end{eqnarray}
Only the local $U(1) \times \overline{U(1)}$ out of $\so(4)$ is left when we turn off $2\times 2$ off diagonal blocks of the gauge fields (\ref{gaugeansatz}). Each $A_\mu{}^1{}_2$ and $A_\mu{}^3{}_4$ corresponds to  $U(1)$ and $\overline{U(1)}$ subsequently. As is usually done, we decompose complex valued $\Phi(t, z, \bar z)$ in terms of its magnitude and of its phase as $\Phi = e^{-g + i \varphi}$. By gauge choice we may set $\Lambda^{1}{}_{2}=\mp \varphi(t,z,\bar z)$ so that $\Phi=e^{-g}$, $A_{z}{}^1{}_2 = \pm i \partial_{z}g$  locally. The solutions that were studied in \cite{NWKIM} was for real $\Phi$ without any global phase. Here we are going to consider a solution whose global $U(1)$ phase is non zero such that
\be
 \ba{lll}
\Phi =e^{-g(r) +i N \theta} \,, ~~~ A_{z}{}^1{}_2 =  i {N - a(r) \over 2 z}\, ,
\ea \ee
where $i \varphi(t, z, \bar z) $ is chosen as $i N \theta$ with $z = r e^{i\theta}$.
To make gauge field well defined at the origin one should have $a(0)= N $. At infinity we take the boundary condition $a(\infty) = 0 $ to make $D_z \Phi$ vanish asymptotically so as to achieve a finite energy system. And that implies quantized magnetic flux over the whole M2 since $\oint_{r\rightarrow \infty } {A}_{i}{}^{1}{}_{2} ~dx^i = - 2 \pi N$. It is worth to note that in the Bagger-Lambert theory of $\so(4)$ gauge group representation, the magnetic flux itself $F_{z\bar z}{}^a{}_b$ is not physical since it clearly breaks $\so(4)$ invariance. Moreover the energy is not bounded by magnetic flux unlike usual Chern-Simons vortices. Since we focus on the abelian nature of the theory and on the construction of a localized object, we apply the same well known properties of the vortex so as to make the flux for the very unbroken $U(1)$ be quantized. However the vortex equation (\ref{vortexeqn}) specifies the relation between the magnitude of $\Phi$ and the unbroken $U(1)$ gauge field. Plugging this relation into the Gauss constraint equation we can get the ordinary equation to specify the whole profile of the vortex. In addition it is worth to note that $F_{xy}{}^3{}_4 = 0$ since all  $D_t X_I = 0$ except $I=3\,,4$. Therefore the field strength of $\overline{U(1)}$ gauge potential should vanish everywhere on the $M2$ world volume and $ A_i{}^3{}_4 $ should be a pure gauge. We can also settle down regularity issue conveniently when $\det M = m$,
\be\ba{lll}
   \tilde{F}_{xy}{}^1{}_2 &= &- 2 \sum |M_i{}^j|^2 (\Phi \bar \Phi  - {m^2 \over \sum |M_i{}^j|^2}) \, , \\

  R_{34} &= & 2m^2 (  \sum |M_i{}^j|^2 -  4 \Phi \bar \Phi  ) \, .  \label{fluxandrotation}

\ea\ee
Solving vanishing magnetic flux and angular momentum at infinity $F_{xy}{}^1{}_2(\infty) = R_{34}(\infty) =0$, we obtain asymptotic values of fields as  $2 \Phi \bar \Phi = m $ and ${1\over 2} \sum |M_i{}^j|^2 = m$. These boundary values automatically set  $V(\infty) = 0$ in (\ref{potential}). For energy finiteness it is an important property but is not unexpected because we have seen that the energy is bounded below by the angular momentum $R_{34}$ and the central charge $Z_{12}$ which consists of magnetic flux of $U(1)$ gauge potential multiplied by $ X^I $s and the other covariant derivative terms (\ref{energybound}).


\subsection{Static $M5$ }\label{sectionvortexsolution}

So far (\ref{threealgebrarelation}) and the right side equations to the semicolons in all (\ref{timedependent}), (\ref{covariantlyconstant}) and (\ref{threealgebrarelation}) have been solved. We are going to solve each of the remaining equations in (\ref{timedependent}), (\ref{covariantlyconstant}) subsequently, leaving the holomorphic covariance (\ref{holomorphiccovariance}) and  the Gauss constraint.  Denote $M_i{}^j$ as,
\be\ba{ll}

M = \pmatrix{ X & -Y \cr
		Y & X
} \,.
\ea\ee
BPS equations (\ref{covariantlyconstant}) for $X_3$ and $X_4$ can be solved as 
$A_i{}^3{}_4 = {\partial_i X \over Y} = - {\partial_i Y \over X}$. Hence $ \partial_i (X^2 + Y^2) = 0 $. 
The condition (\ref{simplicityM}) fixes integration constant and therefore it is natural to write
$  X= \sqrt m \cos(\zeta(x,y) + w t) $ and  $  Y= \sqrt m \sin(\zeta(x,y) + w t) $. Then $  A_i{}^3{}_4 = -\partial_i \zeta $ which is a pure gauge and is consistent with the fact that magnetic flux $F_{xy}{}^3{}_4$ is zero everywhere. The frequency $w$ is fixed by equations (\ref{timedependent}),
\be\ba{lll}
& D_t X_4 = -[X_4, X_1, X_2] - m X_3 \, ,  \\
& D_t X_3 = -[X_3, X_1, X_2] + m X_4  \, ,
\ea\ee
which are simply equations for oscillators. $\ddot{X} = - w^2 X$ and $\ddot{Y} =- w^2 Y$ where $w = 2 \Phi \bar\Phi - A_t{}^3{}_4 -m$. These imply the equations of motion for the electric field automatically and it further restricts $A_t{}^3{}_4$ by,
\be\ba{lll}
\tilde{F}_{tz} X_4 &=& ( D_z D_t - D_t D_z )X_4 = -D_z [X_4, X_1, X_2] - m D_z  X_3 \, ,  \\
 &= &  i D_z [X_4, X_\omega , X_{\bar \omega}] \, , \\
\tilde{F}_{tz} X_3 &=&   i D_z [X_3, X_\omega , X_{\bar \omega}] \, . \\
\ea\ee
Implementing ansatz and using BPS equations again, $\partial_z A_t{}^3{}_4 = \partial_z ( 2 \Phi \bar \Phi )$ together with complex conjugate of the equation, we determine $  A_t{}^3{}_4 = 2 \Phi \bar \Phi - C$. The frequency $w$ is arbitrary up to integration constant.  In fact this constant is a gauge degree of freedom and can be chosen to be $m$ so as to make $X$ and $Y$ be static. Therefore $c = m $. But even though $X$ and $Y$ are static, we have non-vanishing M-wave frequency so is the angular momentum on 3,4 plane because $A_t{}^3{}_4$ carries it.
Assuming $X_1$ and $X_2$ be time independent, $A_t{}^1{}_2 = 0$ when the equations $ D_t X_1 = 0 \, , D_t X_2 = 0 $ in (\ref{timedependent}) are solved. The  solution which solves all the equations in 
 (\ref{timedependent}),  (\ref{covariantlyconstant}), (\ref{threealgebrarelation}) is summarized as,
\be
 \ba{lll}
 \left( \ba{cccc} \vec{X}_\omega   & \\ \vec{X}_{\bar \omega} & \\ \vec{X}_3 &\\ \vec{X}_4 & \ea \right)
 = \left(\ba{cccc}\Phi & i\Phi &0&0\\ \bar \Phi &-i\bar \Phi &0&0\\
 0&0& \sqrt m \cos(\zeta)&-\sqrt m \sin(\zeta)\\0&0&\sqrt m \sin(\zeta)&\sqrt m \cos(\zeta) \ea\right)\,, \label{solutionX}
\ea
 \ee
\be\ba{ll}
\tilde{A}_{z}=\left(\ba{cccc}0& i {N - a(r) \over 2 z}&0&0\\ - i {N - a(r) \over 2 z}&0&0&0\\
 0&0&0&-\partial_i \zeta \\0&0& \partial_i \zeta &0 \ea\right)\,,&~
  \tilde{A}_{t}=\left(\ba{cccc}0&0&0&0\\ 0&0&0&0\\
 0&0&0& 2 \Phi \bar \Phi - m\\0&0& -2 \Phi \bar \Phi +m &0\ea \right)\,. \label{vortexsolution}
 \ea
 \ee
Undetermined real function $ \zeta $ can be gauged away. We are left with the equation (\ref{holomorphiccovariance}) and the Gauss law, which give the well known vortex equations. They are 
\be\ba{ll}
D_z \bar \Phi = 0 \, , \\
\tilde{F}_{xy}{}^1{}_2 =  - 4m( \Phi \bar \Phi - {m\over 2})\, , \label{vortexequationsARCHAIC}
\ea\ee
with the boundary conditions  $F_{xy}{}^1{}_2 (\infty) = 0$ and $R_{34} (\infty) = 0$ which set $|\Phi | $ go to $ \sqrt{m \over 2}$ at the boundary of the $M2$ world volume. As defined, $\Phi = e^{-g(r)+i N \theta}$. $g$ and $a$ are determined by the ordinary equation under the boundary conditions  specified in (\ref{fluxandrotation}), which are $a(0) = N$, $a(\infty) = 0$ where we can win a regular profile.  The covariantly holomorphic equation determines the magnitude of $\Phi$ in terms of $a(r)$ $\it i,e$ $e^{-g} = \sqrt {m \over 2} e^{\int_r^\infty {a\over r'} dr'}$ that is indeed set to $\sqrt {m \over 2}$ asymptotically. The ordinary equation for $a$ is read from the Gauss constraint which is $ a^{''} - {2 a + 1 \over r}a' - 4 m^2 a = 0 $. As is well known it is a nonlinear equation and can be solved numerically. The corresponding behavior of $e^{-g}$ near the origin is regular and has zero value at the origin when we numerically plot the curve \cite{MirrorSymmetry}\cite{abelianvortexCS}. The energy is bounded below by the angular momentum on $3,4$ plane only because covariant derivative on Higgs field vanishes asymptotically for the vortex solutions so that  $\int d^2 x  Z_{12} = 0$. Moreover it is quantized : $\mathcal E \geq \int d^2 x R_{34} =- 8 m^2  \int d^2 x  (\Phi \bar \Phi - {m\over 2}) = 2 m \int d^2 x \tilde{F}_{xy}{}^1{}_2 = 2 \cdot 2m \pi N$. With the gauge choice of $\zeta(x,y)=0$,  the excitations on $X_1, X_2, X_3, X_4$ are asymptotically : 
\be
 \ba{lll}
 \left( \ba{cccc} \vec{X_1}   & \\ \vec{X_2} & \\ \vec{X_3} &\\ \vec{X_4} & \ea \right)
 = \sqrt m \left(\ba{cccc} \cos{N\theta} &\ - \sin{N\theta} &0&0\\ -\sin{N\theta}& -\cos{N\theta}&0&0\\
 0&0&1 &0\\0&0&0&1 \ea\right)\,.\ea
 \ee

%

\section{Discussion}

An incorporation of $U(1)$ vortex in Chern-Simons-Higgs theory of $\so(4)$ gauge group has been performed. Consequently a regular finite energy profile of $M2$ excitation is shown to exist in the ${ 1 \over 2 }$ BPS configuration of  the mass deformed Bagger-Lamber theory with $\so(2) \times \so(2)\times \so(2) \times \so(4)$ global symmetry. All the explicit supersymmetric configurations for the BLG theory that have been presented in \cite{NWKIM} were singular and it was partly because the BLG theory lacks a dimensionfull
parameter which could set out the regularization scale. It was expected that we might have a regular ${1\over 2}$ BPS object in the mass-deformed Bagger-Lambert
theory, and we have shown that indeed we have such an object in the mass deformed theory. ${1\over 4}$ BPS objects like q-balls and vortices are discussed in \cite{massdeformBL} and such ${1\over 4}$ BPS configurations may correspond to some bound state of these ${1\over 2}$ BPS object. Non Abelian vortex \cite{nonabelianCS:Schaposnik} \cite{nonabelianCS:Schaposnik2007} may exist in the BLG theory without mass parameter and it will be interesting object, but it is somehow difficult to find. Therefore we concentrated on the $U(1)$ vortices only.
We reiterate our result in a concise way. Since we have taken $\so(4)$ representation and turned on four coordinates only, we may put $X \equiv X_A{}^a$ as $4 \times 4$ matrix, where $A$ spans $1,2,3,4 $.  We dropped $2 \times 2$ off diagonal blocks both for Higgs field and for the gauge fields to manifest the $U(1) \times \bar U(1)$ unbroken gauge symmetry. By setting $\zeta(x,y)=0$,
\be
X = \left [ \matrix{  \pmatrix{ \Phi   & i\Phi \cr \bar\Phi &  -i\bar\Phi }     & 0 \cr
				0 & \sqrt m \bf 1
					 } \right], 
 \tilde{A}_{z}=\left[\matrix{  \pmatrix{ 0   & A_z{}^1{}_2 \cr -A_z{}^1{}_2 & 0 }     & 0 \cr
				0 & 0
					 } \right], 
  \tilde{A}_{t}=\left[\matrix{  0     & 0 \cr
				0 & \pmatrix{ 0   & 2 \Phi \bar \Phi - m  \cr -\Phi \bar \Phi + m   & 0 }
					 } \right]. 
\ee
where we see a vortex (\ref{vortexsolution}) on $X_1$ and $X_2$ and a constant excitation on $X_3$ and $X_4$. The function $\Phi$ together with $A_z{}^1{}_2$  satisfies the well known Abelian vortex equations(\ref{vortexequationsARCHAIC}). The nonvanishing charges associated with each $2 \times 2$ diagonal excitations are schematically, 
\begin{eqnarray}
~~~~~~~~~~~~~~~~ Magnetic ~~~~~~~~~~~~~~~~~~~~~~~~~~Electric~~~~~~~~~~~~~~~~~~Angular~Momentum~~~~~~\nn \\
 \left [ \matrix{ \int d^2 x ~\tilde{F}_{xy}{}^1{}_2=2 \pi N    & 0 \cr
				0 & 0
					 } \right] \,, ~~~
			\left [ \matrix{ 0   & 0 \cr
				0 &  \int d^2 x ~ \tilde{F}_{0i}{}^3{}_4
					 } \right]
					 \,, ~~~
			\left [ \matrix{ 0   & 0 \cr
				0 &  \int d^2 x ~R_{34} = 4 m \pi N
					 } \right]\,. \nn
\end{eqnarray}
The M-wave along $X_3$ and $X_4$ has angular momentum coming solely from the gauge potential $\tilde{A}_t{}^3{}_4$, and the angular momentum is related to the electric charge as a consequence. The M-theory configuration for MW-M2-M2-M5 bound state can be summarized in table.
\begin{table}[htb]
\begin{center}
\begin{tabular}{lccccccccccc}
   M2: & t & x & y & - & - & - & - & - & - & - & - \\
   M2: & t & - & - & 1 & 2 & - & - & - & - & - & - \\
   MW: & t & - & - & - & - & 3 & - & - & - & - & - \\
 	& t & - & - & - & - & - & 4 & - & - & - & - \\
   M5: & t & x & y & 1 & 2 & 3 & - & - & - & - & - \\
	 & t & x & y & 1 & 2 & - & 4 & - & - & - & - \\
\end{tabular}
\label{m2:m2:kk:mw}
\end{center}
\end{table}

The limit $m\rightarrow 0$ is trivial because the magnetic flux vanishes (\ref{vortexequationsARCHAIC}). Multi-vortex solution is straightforward. The vortex scattering problem will be an interesting topic. Switching our result into $\su(2)\times \su(2)$ representation which is identical to $\so(4)$ we used can readily be performed \cite{Raamsdonk}. The ABJM \cite{ABJM} generalization of this object and the gravity duals in AdS/CFT perspective might be other ways for further study and they are partly under progress.
\\
\\
\\
${\bf Acknowledgement}$
\\
\\
J. K. would like to thank Ki-Myoung Lee, Jeong-Hyuck Park and Masato Arai for insightful discussions and advice. This work is supported by the Center
for Quantum Spacetime of Sogang University with grant number R11 - 2005 - 021.

\end{document}